\documentclass[twocolumn,times]{aastex61}
\pdfoutput = 1
\usepackage{underscore}
\usepackage{float}
\usepackage[caption = false]{subfig}
\usepackage{graphicx}
\usepackage[T1]{fontenc}
\usepackage[bookmarks=false]{hyperref}
\usepackage{afterpage}

\journalinfo{Accepted to The Astronomical Journal}

\shorttitle{Activity \& Kinematics of SUPERBLINK WD+dMs}

\begin{document}

\title{Activity and Kinematics of White Dwarf-M Dwarf Binaries from the SUPERBLINK Proper Motion Survey\footnotemark[1]}
\footnotetext[1]{Based on Observations obtained at the MDM Observatory operated by Dartmouth College, Columbia University, The Ohio State University, and the University of Michigan}

\author{Julie N. Skinner}
\affiliation{Institute for Astrophysical Research, Boston University, 725 Commonwealth Avenue, Boston, MA 02215, USA}

\author{Dylan P. Morgan}
\affiliation{Department of Astronomy, Boston University, 725 Commonwealth Avenue, Boston, MA 02215, USA}

\author{Andrew A. West}
\affiliation{Department of Astronomy, Boston University, 725 Commonwealth Avenue, Boston, MA 02215, USA}

\author{S\'ebastien L\'epine}
\affiliation{Department of Physics \& Astronomy, Georgia State University, 25 Park Place NE, Atlanta, GA, 30303, USA}

\author{John R. Thorstensen}
\affiliation{Department of Physics and Astronomy, 6127 Wilder Laboratory, Dartmouth College, Hanover, NH 03755, USA}

\correspondingauthor{Julie N. Skinner}
\email{jskinner@bu.edu}

\begin{abstract}
We present an activity and kinematic analysis of high proper motion white dwarf-M dwarf binaries (WD+dMs) found in the SUPERBLINK survey, 178 of which are new identifications.  To identify WD+dMs, we developed a UV-optical-IR color criterion and conducted a spectroscopic survey to confirm each candidate binary.  For the newly identified systems, we fit the two components using model white dwarf spectra and M dwarf template spectra to determine physical parameters.  We use H$\alpha$ chromospheric emission to examine the magnetic activity of the M dwarf in each system, and investigate how its activity is affected by the presence of a white dwarf companion.  We find that the fraction of WD+dM binaries with active M dwarfs is significantly higher than their single M dwarf counterparts at early and mid spectral types. We corroborate previous studies that find high activity fractions at both close and intermediate separations.  At more distant separations the binary fraction appears to approach the activity fraction for single M dwarfs. Using derived radial velocities and the proper motions, we calculate 3D space velocities for the WD+dMs in SUPERBLINK.  For the entire SUPERBLINK WD+dMs, we find a large vertical velocity dispersion, indicating a dynamically hotter population compared to high proper motion samples of single M dwarfs.  We compare the kinematics for systems with active M dwarfs and those with inactive M dwarfs, and find signatures of asymmetric drift in the inactive sample, indicating that they are drawn from an older population.
\end{abstract}

\section{Introduction}
\label{sec:intro}
Low-mass stars are the most populous stellar constituents in our galaxy and are vital to understanding our galaxy's structure and kinematics \citep{Gizis02, West04, Bochanski07b}.  Over a quarter of low-mass stars are found in binary or higher-order systems \citep{Duchene13}, making it crucial to understand how binary companions affect low-mass star formation and evolutionary processes.   One such process is the evolution of stellar magnetic fields with age and the subsequent change in the magnetic heating of the upper atmosphere. Despite many observations of magnetic activity, the exact mechanism behind chromospheric and coronal heating in low-mass star atmospheres is not well-understood.  In solar-type stars, differential rotation between the inner radiative zone and outer convective layer causes rotational shear that organizes and amplifies the magnetic field, with faster rotating stars having stronger magnetic fields.  This strong magnetic field provides the energy to heat the stellar chromosphere and corona and gives rise to line emission (as well as continuum), which is used to trace the magnetic activity.  Over time, the stellar rotation period decreases due to loss of angular momentum from the magnetized winds leaving the star, linking stellar rotation, age, and magnetic activity.

Previous studies have found that the rotation-activity correlations observed in solar-type stars also apply to early-type M dwarfs (M0-M3; \citealt[e.g.][]{Kiraga07}).   However, around spectral type $\sim$M4, M dwarfs transition to a fully convective interior (0.35 M$_{\odot}$; \citealt{Chabrier97}) and rotational shear at the tachocline can no longer serve as an explanation for the creation of large scale magnetic fields.  Late-type M dwarfs show the observational signatures of strong magnetic activity \citep{Delfosse98,Mohanty03,Reiners07}, but the full details of the relationship between fully convective stars, rotation, age, and the generation of the magnetic field are not yet fully understood.   However, recent work from \citet{West15} showed that the high activity fraction observed in a sample of late-type M dwarfs persists to shorter rotation periods than for early-type M dwarfs, indicating that rotation is important in these late-type stars and the activity-rotation relation is mass dependent.

Binaries provide a window into the physical characteristics of stars not usually afforded from their single counterparts.  Close, interacting systems allow us to better understand accretion properties and mass loss through winds.  Wide binary systems with M dwarfs are often used to better understand the abundances and ages of low-mass stars, through a study of their co-eval companions.  In addition, if these systems are eclipsing, complete orbital properties can be derived and therefore fundamental stellar parameters can be determined.  Eclipsing binaries often act as anchor points for stellar evolution models \citep[][and references therein]{Feiden15}, highlighting the importance of understanding at which separations interactions between the stars affect the observed and physical properties of the components.  The previous categories of close or wide, while useful, ignore intermediate separations where interaction between the stars may be more subtle.  In this work, we will highlight the need to use three categories of binary separation\footnote{These regimes are chosen based on the orbital period distribution of the population of white dwarf-main sequence binaries in the Galaxy as determined by population synthesis models \citep[see Figure 10][]{Willems04}.} -- close (a $\leq$ 0.88 AU), intermediate (0.88 AU < a $\leq$ 2.57 AU), and wide (a > 2.57 AU).  

%\footnotetext[1]{These regimes are chosen based on the orbital period distribution of the population of white dwarf-main sequence binaries in the Galaxy as determined by population synthesis models \citep[see Figure 10][]{Willems04}.}

White dwarf-M dwarf binaries (WD+dMs) provide an important laboratory to explore how close companions affect M dwarfs.  One advantage of studying WD+dMs is the distinct spectral signature from each star.  White dwarfs and M dwarfs have similar luminosities, but differing SEDs, making separation of each component relatively easy.  One avenue to examine is how magnetic activity changes due to binary companions.  To fully understand the magnetic activity evolution of WD+dMs, it is important to have large, statistically significant samples of these binaries. The initial samples of wide, common proper motion white dwarf-main sequence binaries (WDMS) were identified by \citet{Luyten97}. In recent years, there have been several studies of WDMS that have primarily been identified from the Sloan Digital Sky Survey (SDSS; \citealt{York00}). \citet{RM16} presented the updated SDSS DR1-12 spectroscopic catalog of WDMS binaries with 3294 systems.  The same team previously employed a two step color selection using both SDSS \textit{ugriz} photometry, as well as a search for infrared excess in the Two Micron All Sky Survey (2MASS; \citealt{2mass}, the Wide-field Infrared Survey Explorer (\textit{WISE}; \citealt{Wright10}),  and the UK Infrared Telescope (UKIRT) Infrared Sky Survey \citep{Lawrence07,Warren07,RebassaMansergas13}.  Other studies utilizing the SDSS catalog include the \citet{Silvestri06, Heller09, Morgan12, Liu12}.  Additionally, 121 WDMS binaries were identified from the Large sky Area Multi-Object Spectroscopic Telescope (LAMOST) DR1 \citep{Ren14}.  While all of these studies have significant overlap, \citet{RM16} presents the largest, homogeneous sample of spectroscopically confirmed WDMS systems.  Prior to these larger studies, \citet{Raymond03} identified $\simeq$ 100 WD+dMs from SDSS and \citet{Schreiber03} presented a listing of 30 known post-common envelope binaries (PCEBs) identified through a variety of methods.  In most cases, the WD+dMs in these studies have been selected from SDSS, and are biased against cool white dwarfs or early-type M dwarfs \citet{RebassaMansergas13}.  WD+dMs from SDSS have spectroscopic distances typically greater than a few hundred parsecs, leaving many of the solar neighborhood WD+dMs unidentified.  Samples of more nearby WD+dMs, on the other hand, offer the opportunity to study the kinematics of these binaries, as well as the opportunity to build a volume-limited sample of WD+dMs.  Surveys of high proper motion stars, like the SUPERBLINK proper motion survey \citep[e.g.][]{Lepine11} make it possible to examine more nearby WD+dMs.

Previous studies examining the activity of M dwarfs with white dwarf companions \citep{Silvestri07,Morgan12,RebassaMansergas13a} have shown that the fraction of active early- and mid-type M dwarfs with white dwarf companions is higher than their single counterparts.  For late-type M dwarfs, activity fractions for WD+dMs approach similar levels as single M dwarfs, suggesting that angular momentum loss in fully convective single M dwarfs is less efficient than for early and mid-types.  \citet{Silvestri05} found activity fractions for early- and mid-type M dwarfs with wide white dwarf companions that were similar to the single M dwarf activity fractions.  However, \citet{Morgan12} and \citet{RebassaMansergas13a} found that WD+dMs at intermediate-to-wide separations have higher activity fractions than similar samples of single M dwarfs.  \citet{RebassaMansergas13a} attribute this high activity fraction to the preferential selection of systems with younger, hotter white dwarfs.  

Since binaries at different separations inhabit different dynamical regimes, the physical mechanism producing higher magnetic activity fractions in WD+dMs likely varies as a function of binary separation.  The closest WD+dMs (P$_{orb}$ $\lesssim$ 10 days, \citealt{GomezMoran11}) are products of common envelope evolution, the binary evolution state in which the envelope of the WD progenitor engulfs the companion M dwarf.  Binary angular momentum is subsequently extracted to eject the envelope, and a very close WD+dM emerges as a post-common envelope binary or PCEB. In these systems, tidal locking ensures that the M dwarf will be rapidly rotating, and therefore likely observed as active.  \citet{RebassaMansergas13a} found that all of the confirmed PCEBs in their sample were active across all observed M dwarf subtypes (M2-M7).  \citet{Morgan12} also found that the closest WD+dMs (a < 0.1 AU) have high activity fractions across all M dwarf subtypes, while the widest systems (a = 1-100 AU) had low activity fractions at early M spectral types with an increasing activity fraction toward late spectral types.

WD+dMs at wider separations do not undergo common envelope evolution, and are often assumed to evolve as single stars.  However, differences in the lifetime of the circumstellar disk at early epochs may play a significant role in the subsequent angular momentum evolution.  To avoid approaching rotation speeds close to the breakup limit, single stars must shed the angular momentum gained during the pre-main-sequence stage.  Single stars may dissipate angular momentum through accretion-powered winds \citep{Matt05b}, or through disk locking \citep{Ghosh79, Shu94}, which applies a torque on the star through a magnetic bridge between the star and the disk.  In binaries, the companion star may disrupt the disk, and thus shorten the time available for either mechanism to carry away angular momentum, resulting in a system that arrives at its post-disk life with a faster rotation period.  \citet{Meibom07} found that the solar-type primary stars of binaries (0.1 AU $\lesssim$ a $\lesssim$ 5 AU) in M35 had faster rotation periods on average than solar-type single stars in the same cluster.  The effects disk disruption may have at later times are crucial to understanding the separation threshold at which stars in binaries can be assumed to evolve as single stars (i.e. this assumption may not hold for stars at intermediate separations).

In this paper, we present the sample of WD+dM binaries extracted from the SUPERBLINK proper motion survey. In section \ref{sec:targets}, we outline our target and color selection, as well as spectroscopic observations from our subsequent, follow-up program.  Section \ref{sec:analysis} outlines our two-component fits to the recorded spectra, as well as the resulting spectroscopic distances and orbital characteristics that we derive from these fits.  In section \ref{sec:results}, we present the 3D kinematics of this sample and discuss the results of our magnetic activity analysis.  Lastly, a summary of the results and concluding remarks are presented in section \ref{sec:conclusion}. 

\begin{figure}
\figurenum{1}
\includegraphics[angle= 0,width={\columnwidth}]{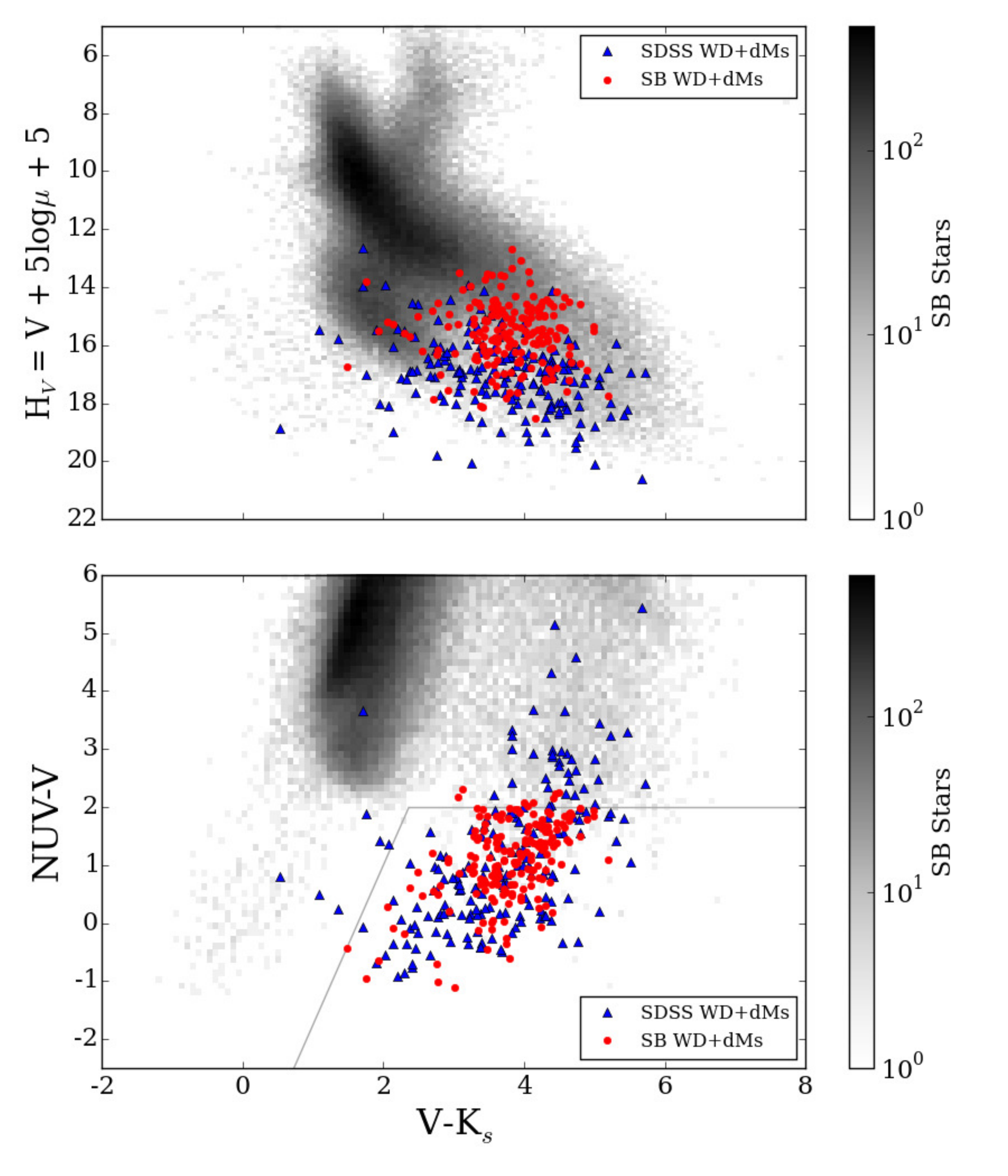}
\caption{Top Panel: The reduced proper motion diagram for SDSS WD+dMs found in SUPERBLINK \citep[blue triangles][]{RebassaMansergas10} and the SUPERBLINK WD+dMs presented in this work (red dots) alongside the main stellar locus from SUPERBLINK (in gray).  Bottom Panel: The \textit{NUV-V} and \textit{V-K$_{s}$} color-color diagram is shown with the SUPERBLINK stars plotted in gray.  WD+dMs identified in this study (red dots) and WD+dMs previously identified in SDSS (blue triangles) have significant excesses in the near-UV and infrared K.  Gray lines indicate our initial color-space selection, from which we conducted our spectroscopic survey.  Some previously known WD+dMs fall outside our selection box, indicating that our color selection scheme should be expanded to redder \textit{NUV-V} colors.}
\label{fig:colorselect}
\end{figure}

\section{Target Selection \& Observations}
\label{sec:targets}

To select for nearby WD+dMs, we used the SUPERBLINK proper motion survey \citep{LSPM02,LSPM05}, an ongoing all-sky survey which identifies and characterizes stars with proper motions $\mu >$ 40 mas yr$^{-1}$.  The SUPERBLINK catalog is assembled from an automated search of the Digitized Sky Surveys (DSS), followed by visual examination of objects flagged with large proper motions.  Images from two epochs are aligned and normalized before being differenced to produce a residual image; objects with large proper motions are identified as high intensity pairs on the residual image by the software, and visually checked by blinking the finding charts generated by SUPERBLINK.  For this study, we used the July 2011 version of SUPERBLINK, which listed 2,270,481 stars, and was estimated to be $>$90 \% complete to \textit{V} = 19.0.  The catalog is known to be more complete at higher galactic latitudes ($>$95\%) where crowding is minimal, and the low-latitude completeness is estimated to be $>$80-85\%  \citep[][and references therein]{Lepine11}.  SUPERBLINK has been cross-matched with photometric archives including the \textit{ROSAT} All-Sky Survey Bright Source Catalog \citep{Voges99}, the \textit{ROSAT} All-Sky Survey Faint Source Catalog \citep{Voges00}, the fifth data release from \textit{Galaxy Evolution Explorer} (\textit{GALEX}; \citealt{Martin05}), and the Two Micron All Sky Survey (2MASS; \citealt{2mass}) catalog.  The astrometric and photometric properties for the SUPERBLINK sample of WD+dMs are shown in Table \ref{tab:astrophot} and a reduced proper motion diagram with the SUPERBLINK WD+dMs is shown in the top panel of Figure \ref{fig:colorselect}. As expected, the WD+dM pairs straddle the main sequence and white dwarf cooling sequence in the reduced proper motion diagram. None of the objects in our sample had X-ray detections from ROSAT, so we have not included those data in the table.

We selected WD+dMs based on a combination of \textit{V} magnitudes derived from the DSS plates (see \citealt{LSPM05}), near-\textit{UV} magnitudes from \textit{GALEX}, and \textit{K$_{s}$} magnitudes from 2MASS.  Using the UV-optical-IR color-selection outlined in \citet{Skinner14}, we selected targets for spectroscopic follow-up (see bottom panel of Figure \ref{fig:colorselect}).  We acquired optical spectroscopy of 178 newly identified WD+dM candidates, with the Boller and Chivens CCD spectrograph (CCDS), using both the Hiltner 2.4m and McGraw-Hill 1.3m telescopes located at MDM Observatory.  The detector is a Loral 1200 x 800 CCD with several gratings available.  For all CCDS spectra, we used the 150 l/mm grating, which is blazed at 4700~\AA.  With the exception of a portion of the April 2011 observing run, we inserted a LG-350 order blocking filter to block second-order contributions to the spectra.  Slit sizes were 1.2-arcsec at the 2.4m and 1.5-arcsec at the 1.3m, which yielded resolutions of 10 \AA~and 8 \AA, respectively.  Observing conditions over the multiple runs were varied.  Objects confirmed to be WD+dM binaries that were originally observed in poor conditions were typically re-observed in better conditions at a later date.

To maximize efficiency at the telescope, spectra were calibrated using comparison lamp spectra taken during twilight to find the shape of the pixel-to-wavelength relation, then shifts to each spectrum were derived from the 5577 \AA~ night sky line to set the zero-point of the calibration.  To confirm that these shifts were applied correctly, the IRAF task \textit{xcsao} was used to cross-correlate the 6100-6700 \AA~region of each spectrum with an average sky spectrum constructed using a similar instrument.  These reductions were done using standard IRAF routines.  In addition to our target spectroscopy, we obtained spectra of flux standard stars and well-known M dwarf spectral type standards.  The former were used for flux calibration, the latter as templates for spectral typing the M dwarfs discovered in WD+dM systems.  In addition, we observed hot O and B stars to correct for telluric absorption.

For many of the WD+dMs that we confirmed from these spectra, we obtained additional optical spectra at longer wavelengths using the 1.3m McGraw-Hill telescope.  For this, we used the Mark III Spectrograph and the MDM detector Nellie.  Nellie is a STIS 2048$^{2}$ thick, frontside illuminated CCD.  In all MkIII spectra, we used a 300 l/mm grism blaze at 6400~\AA\ with a Hoya Y-50 order blocking filter with a 1.52-arcsec slit.  Reductions were done in the same manner as outlined above.  Spectra from CCDS were used for most of the analysis in this paper, but MkIII spectra were used as additional confirmation of M dwarf spectral type when necessary.

\section{Analysis}
\label{sec:analysis}
\subsection{Two-Component Fits}
\label{sec: fits}
\begin{figure*}
\figurenum{2}
\includegraphics[width=\textwidth]{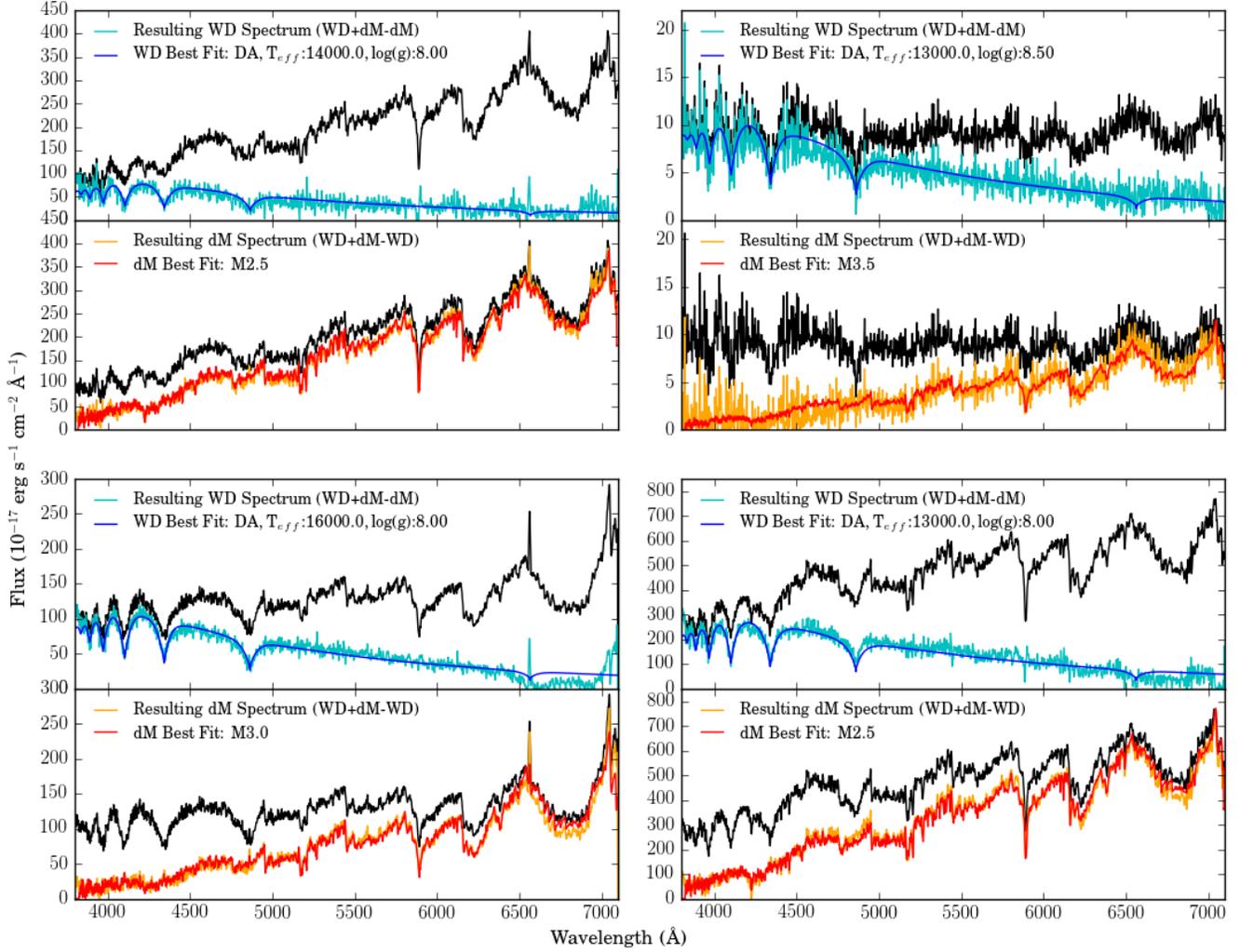}
\caption{Examples of our two-component fitting procedure as outlined in Section \ref{sec: fits}.  Each of the top panels shows our original spectrum in black, the best-fit model WD spectrum in blue, and the target spectrum with the best-fit M dwarf template subtracted in cyan.  Each of the bottom panels shows the original spectrum in black, the best-fit M dwarf template in red, and the target spectrum with the best-fit model WD spectrum subtracted in orange.}
\label{fig:fitexample}
\end{figure*}

\begin{figure*}
\figurenum{3}
\includegraphics[angle= 0,width=\textwidth]{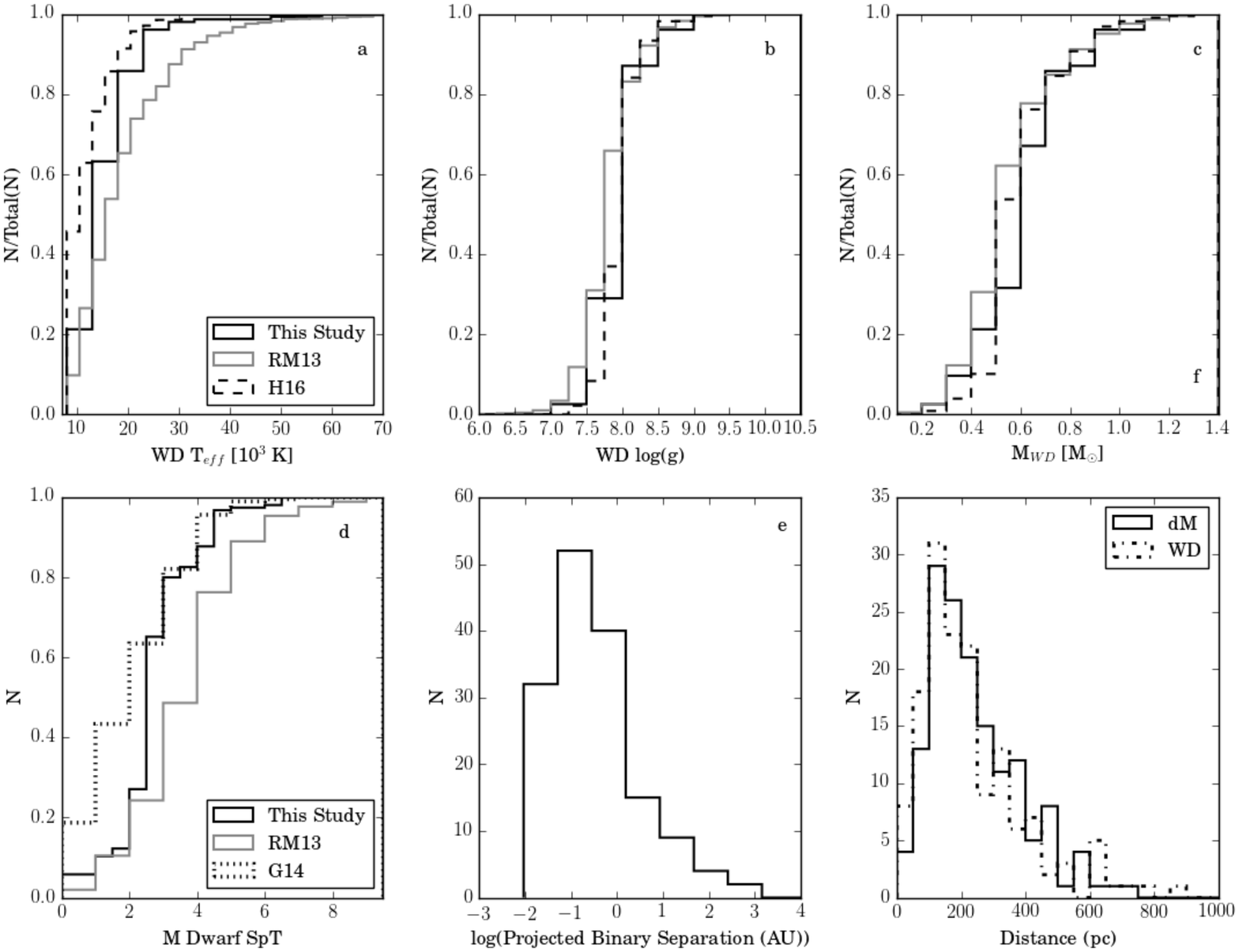}
\caption{Cumulative histograms are shown in panels a-d for WD T$_{eff}$, WD log$g$, M$_{WD}$, and M dwarf spectral subtype (SpT).  Panels e and f show histograms of binary separation and distance, respectively.  In panels a-e, the solid black line shows the newly characterized WD+dMs from SUPERBLINK.  The solid gray line in panels a, b, c, and d shows the WD+dM binaries from \citet{RebassaMansergas13}.  The black dashed line shown in panels a-c represents the volume-limited sample of white dwarfs within 25pc from \citet{Holberg16}.  The black dotted line in panel d represents the sample of SUPERBLINK single M dwarfs from \citet{Gaidos14}.  Panel e shows the distribution in binary separation based on our radial velocity analysis. Panel f shows the spectroscopic distances derived for the M dwarf (solid line) and for the WD (dashdot line).}
\label{fig:cdists}
\end{figure*}

For each of the WD+dM binaries observed, we decomposed each spectrum into the two components with the help of white dwarf model spectra \citep{Koester01a} and M dwarf templates \citet{Bochanski07a}.  Figure \ref{fig:fitexample} shows example spectra with the best-fit model/template combination.  To find the best combination, we followed the iterative $\chi^{2}$-minimization method described in \citet{Morgan12}.  Briefly, the full spectrum is fit first with a white dwarf model that is then subtracted from the spectrum.  An M dwarf template is then fit to the WD-subtracted spectrum.  The process is repeated on the original spectrum, this time fitting the M dwarf template first, then the WD model.  As a part of the fitting process, some spectral regimes (e.g. the Balmer lines from the WD) are given greater or lesser weight according to their contribution to the full spectrum.  Several iterations of this process are performed until the best-fit WD model and best-fit M dwarf template do not change over multiple iterations.  In cases where a solution is not found after 10 iterations, the process is terminated, and the fit with the lowest $\chi^{2}$ value is chosen.  Radial velocities for the both the WD and M dwarf components are determined by cross-correlating the separated spectra with the corresponding best-fit model or template after each iteration.  The radial velocity is then used to obtain a better fit by shifting the templates in subsequent iterations.  The WD model atmosphere grid used in this analysis was coarse, and WD fits to a finer grid of models are warranted at a future time. Figure \ref{fig:cdists} shows the effective temperature, log $g$, WD mass, M dwarf spectral type, binary separation, and distance for the new WD+dMs found in SUPERBLINK, along with comparisons to the appropriate binary and single star samples.  Table \ref{tab:rvsspt} gives the results of our two-component fits, as well as the spectroscopic distances (described below) for the newly identified SUPERBLINK WD+dMs.

\subsection{Spectroscopic Distances}
\label{sec:distances}
Using the physical parameters derived from our spectroscopic fits, we determined spectroscopic distances to each of the components of the binary.  Using the determined effective temperatures and gravities from the WD model fits, we used synthetic color sequences developed by Pierre Bergeron and collaborators\footnote{http://www.astro.umontreal.ca/$\sim$bergeron/CoolingModels/} \citep{Holberg06, Kowalski06, Tremblay11a, Bergeron11}.  Interpolating over their grid for M$_{NUV}$ (P. Bergeron, private communication), we determined absolute near-UV magnitudes for the WDs in each of the systems and then calculated distances.  For single M dwarfs, photometric distance relationships typically have less scatter than spectral type-absolute magnitude relationships \citep{Bochanski11, Lepine13}.  However, for WD+dM systems, the compound colors will not adhere to typical, single M dwarf photometric parallax relations since the white dwarf contributes flux in optical and UV.  To minimize contamination from the WD, we used the spectroscopic parallax relation from \citet{Lepine13} to determine absolute \textit{J-band} magnitudes for the M dwarfs based on their spectral types.  Specifically, the relevant relationships for this study are as follows:
\begin{eqnarray}
 M_{J} = 5.220 + 0.393(SpT) + 0.040(SpT)^{2} \nonumber\\
 		\:for\: Active \:K7-M2.5  \nonumber\\
 M_{J} = 5.680 + 0.393(SpT) + 0.040(SpT)^{2} \nonumber\\ 
 		\:for\: Inactive \:K7-M6 \nonumber
\end{eqnarray}

This relationship reflects the fact that active early-type M dwarfs are systematically more luminous than their non-active counterparts (see \citealt{Bochanski11} and references therein).  We used the early-active relation for stars with subtype M2.5 and earlier.  Uncertainty in the determined M dwarf distances is dominated by the spectral type determination, since the 2MASS \textit{J-band} magnitudes typically have small uncertainties.  As found in \citet{Lepine13}, these uncertainties are typically 25-30\%.  We estimate that the WD distances have uncertainties of $\sim$15\%.  Roughly two-thirds of the systems in our sample have distances that agree within their respective uncertainties.  Our fitting process does not explicitly distinguish between hot and cold WD solutions; a few of the discrepant distances could arise from this degeneracy.  \citet{RebMan07} found a similar result in their study of WD+dMs.  They largely attribute differences in the two distance determinations to spectral types that are too early for the M dwarf mass (due to magnetic activity).  The spectroscopic parallax relation from \citet{Lepine13} does attempt to correct for this in early-type M dwarfs, but it is possible the effect is systematically larger for M dwarfs with close companions.  Additionally, M dwarfs in PCEBs containing hot white dwarfs may show effects of irradiation.

\subsection{Orbit Characteristics}
\label{sec:orbits}
As stated in Section \ref{sec: fits}, we measured radial velocities from our spectra using the cross-correlation of our spectra with the appropriate model or template. White dwarf masses were determined using the photometric calibration models referred to in the previous section.  M dwarf masses were assigned based on spectral type using the relation from \citet{Reid05nlds}.  Each of the measured velocities ($v'_{WD}$,$v'_{dM}$) is the sum of the absolute orbital velocity of that component ($v_{WD}$,$v_{dM}$) and the system velocity ($\gamma$; see \citealt{Morgan12}).  Due to the degeneracy that exists between radial velocity signatures and gravitational redshifts in white dwarfs, we have assumed 32 km s$^{-1}$ offset to the measured white dwarf velocities to account for this gravitational redshift \citep{Falcon10}.  Assuming that the binaries are in circular, edge-on orbits, we can combine the component radial velocities with each stars' mass to derive a lower limit on the orbital radial velocity.  Uncertainties on these velocities were calculated using standard error propagation techniques. We stress that the orbital velocities derived based on these assumptions are lower limits and should only be used for statistical comparison. To estimate each star's projected distance from the system center-of-mass, we assumed each of the components' absolute orbital velocities is its real orbital velocity.  We then summed those values to estimate the total binary separation for each system.  We use these projected separations in a comparative way to investigate how magnetic activity changes with binary separation in Section \ref{sec:magact}.  

\subsection{Individual Objects}
\begin{figure}
\figurenum{4}
\includegraphics[angle=0,width={\columnwidth}]{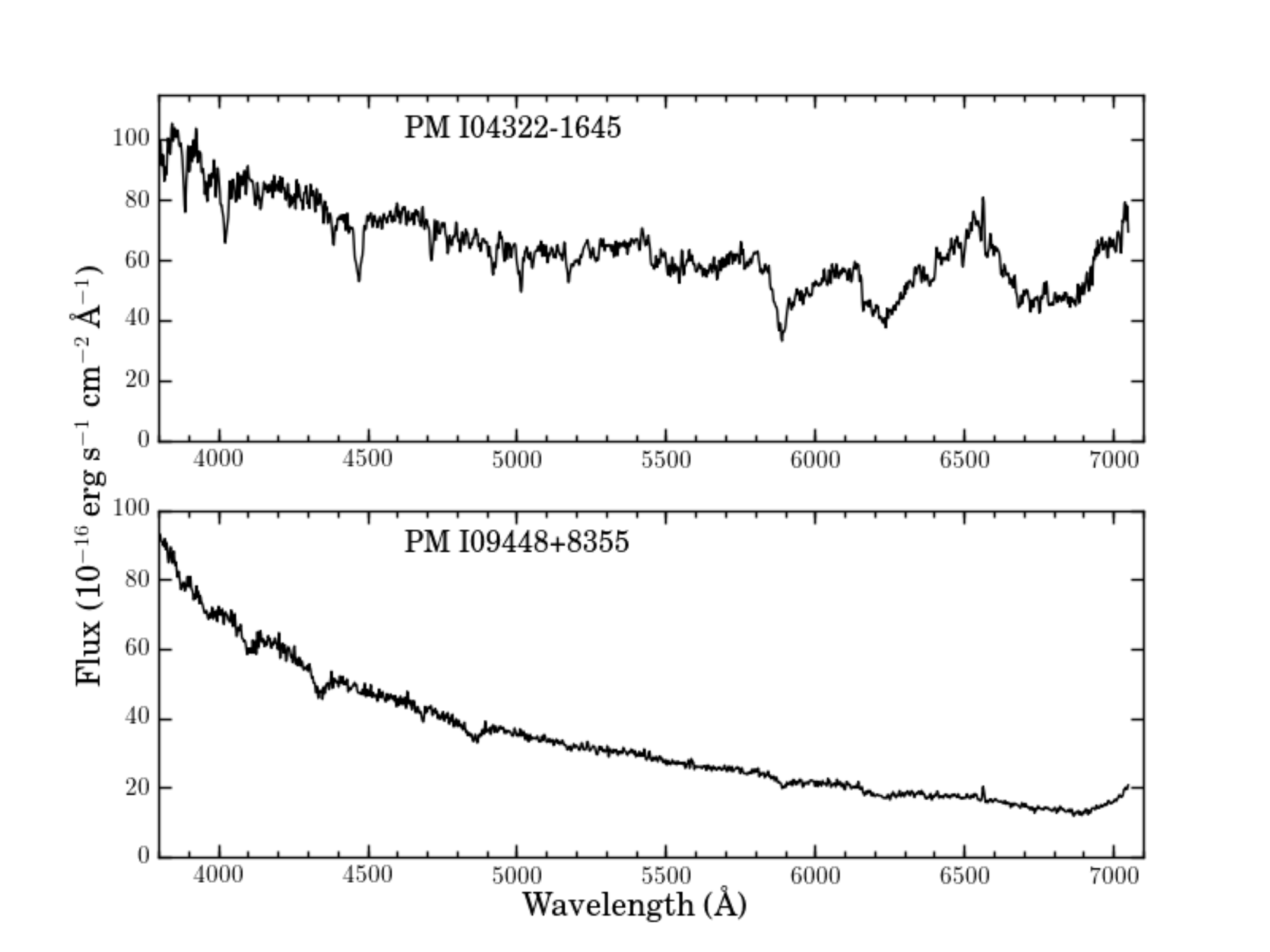}
\caption{Spectra of the two newly identified systems with non-DA white dwarfs.  The top panel shows PM04322-1645, a DB WD previously identified by \citet{Bergeron11}, but no companion was noted.  The bottom panel shows PM09448+8355, a system with a DAO WD identified by this work.}
\label{fig:othertypes}
\end{figure}

In our analysis above, we included only WD+dMs containing a DA WD.  However, we also did identify two systems with non-DA WD spectral classes.  The spectra of these objects are shown in Figure \ref{fig:othertypes}.  The astrometric and photometric properties are included in Table \ref{tab:astrophot}.  PM04322-1645 was previously found by \citet{Bergeron11} as a DB (Helium-line) WD, but there was no companion identified then. We classified PM09448+8355 as a DAO type (mixed H+He composition) WD due to the presence of HeII $\lambda$4686 absorption.

\section{Results}
\label{sec:results}

\subsection{Kinematics}
\label{sec:kinematics}
\begin{figure}
\figurenum{5}
\includegraphics[angle= 0,width={\columnwidth}]{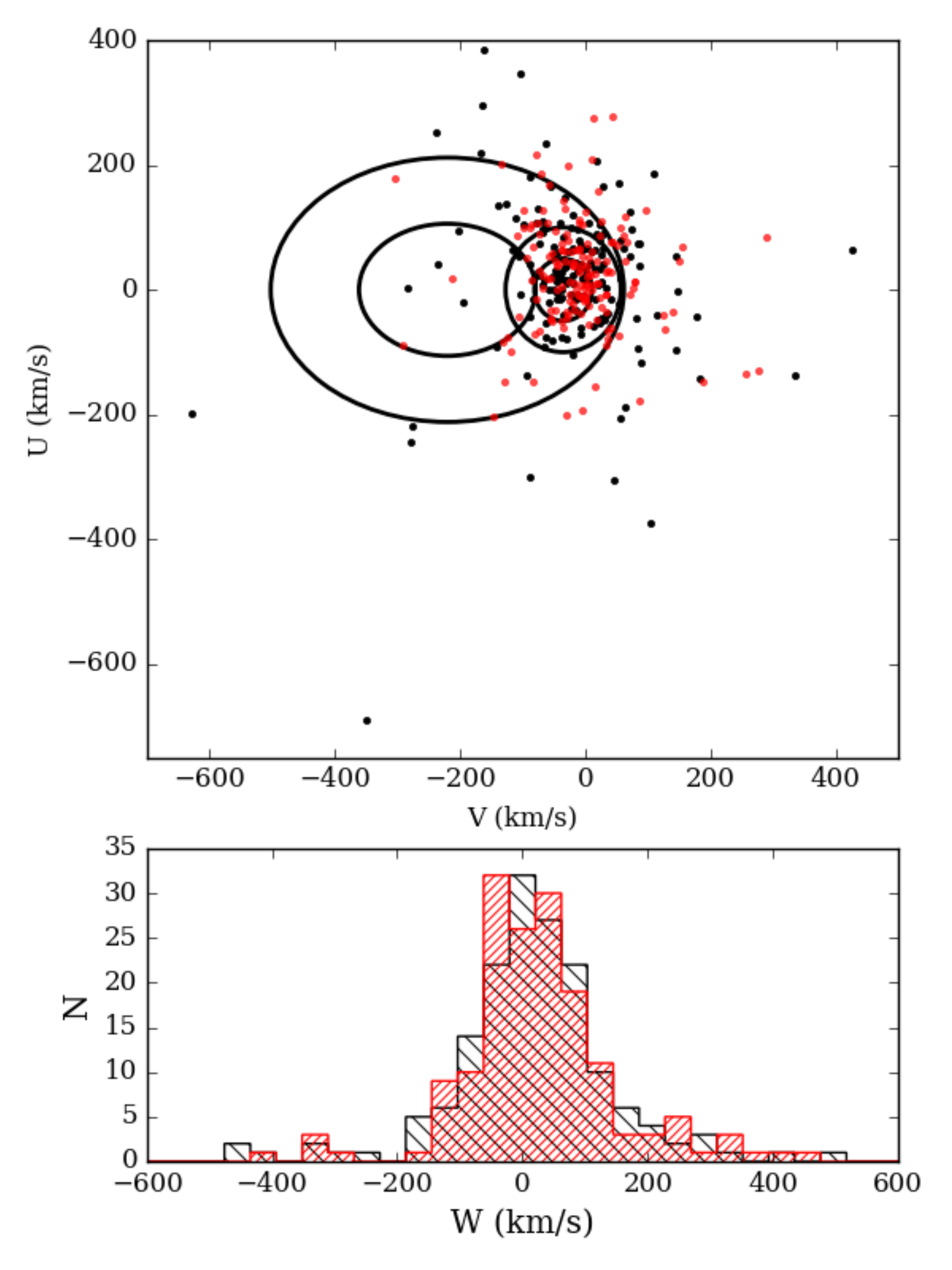}
\caption{UV-velocity distribution for WD+dMs in SUPERBLINK.  Systems with active M dwarfs are shown in red and systems with inactive M dwarfs are shown in black.  Solid black lines denote 1$\sigma$ (inner) and 2$\sigma$ (outer) velocity ellipsoids for stars in the thick disk and halo populations \citep{Chiba00}.  Error bars are omitted for clarity; median uncertainties in $U$ and $V$ are $\simeq$20 km s$^{-1}$.  Bottom Panel:   The $W$ galactic velocity distribution for the SUPERBLINK WD+dMs.  The velocity distribution for active M dwarfs is shown in hatched red overlaying the distribution for the M dwarfs showing no activity in hatched black.  This illustrates the broad distribution in the vertical component of the velocity indicating our sample is drawn from an older population.
}
\label{fig:spacevel}
\end{figure}

Combining the stellar positions, estimated radial velocities, spectroscopic distances determined from our M dwarf fits, and the proper motions from the SUPERBLINK catalog, we derived 3D space velocities for each of the WD+dMs in SUPERBLINK (see Table \ref{tab:kinematics}).  To do this, we made use of the utilities package in \texttt{galpy} \citep{Bovy15}.  We accounted for the solar motion relative to the Local Standard of Rest using values from \citet{Schonrich10} (11, 12, 7 km s$^{-1}$). To determine the formal error for each of the space motions, we calculated the covariance matrix for each ($U,V,W$) vector to propagate the distance, proper motion, and radial velocity uncertainties.  The formal error for each object is listed in Table \ref{tab:kinematics}.  The median uncertainty in each direction ($U,V,W$) is (20, 16, 32 km s$^{-1}$) respectively.  The top panel in figure \ref{fig:spacevel} shows a $U,V$-velocity diagram along with 1$\sigma$ and 2$\sigma$ velocity ellipsoids representing thick disk and halo populations \citep{Chiba00} and the bottom panel of figure shows the distribution of the vertical velocity component of the space motion for the WD+dMs in SUPERBLINK.  Stars classified as active are shown in the red, hatched distribution overlaying the distribution for inactive stars in black hatching.  

Since (for early- to mid-type M dwarfs) active stars should in general be rotating faster and therefore more likely to be from a younger population, we considered three samples for our analysis -- the full sample, only the active stars, and only the inactive stars.  Under the assumption that the underlying velocity distributions for all three kinematic components are Gaussian, probabilities for a given component's mean ($\mu$) and standard deviation ($\sigma$) can be computed analytically and compared to the data using a Bayesian approach.  As described in \citet{West15}, we constructed a grid of $\mu$ and $\sigma$ values and explored the joint probability distribution for the three different subsets of the data: 1) all pairs; 2) pairs with an active M dwarf; and 3) pairs with an inactive M dwarf.  From the resulting peaks of the joint probability density distributions, we computed $\mu$ and $\sigma$ values for each sub-sample.  We also calculated uncertainties from the spread of the marginalized distributions for each parameter. The results of this analysis can be found in Table \ref{tab:popkinematics}.  

In particular, we find that systems with inactive M dwarfs have preferentially lower $V$ (azimuthal) velocities, suggesting that the inactive sample is drawn from an older population exhibiting asymmetric drift.  This is reflected in the lower $\mu_{V}$ for the active and inactive distributions.  We performed a two-sample Kolmogorov-Smirnov test to compare the active and inactive distributions of $V$ velocities and found a $<$ 5\% chance of being drawn from the same population.  This is consistent with results from previous studies \citep{Hawley96,Bochanski07b}, which show that samples of inactive single M dwarfs have more negative $V$ velocities than their active counterparts.  In addition, previous studies of WDs and WD+dMs have investigated the nature of high velocity systems\footnote{$[U^{2} + (V+35)^{2}]^{1/2}$ $>$ 94 km s$^{-1}$} \citep{Silvestri02,Oppenheimer01a}.  The $U,V$-velocity diagram in figure \ref{fig:spacevel} reveals a number of high velocity systems in our sample. More investigation is needed to determine membership in a particular galactic component for individual systems in our sample (e.g. M dwarf metallicity determinations).

We also note that the velocity dispersions in the vertical direction ($W$) for both active and inactive samples are larger than the velocity dispersions typically associated with the thick disk.  \citet{Bochanski07b} found thick disk velocity dispersions ($\sigma_{U},\sigma_{V},\sigma_{W}$) $\approx$ (50 km s$^{-1}$, 40 km s$^{-1}$, 40 km s$^{-1}$ for K and M dwarfs in SDSS.  We attribute this high vertical velocity dispersion to two factors: 1) Selecting M dwarfs with white dwarf companions inherently creates a minimum age limit for our sample since each system must be as old as the progenitor of the white dwarf; and 2) We have preferentially chosen the fast moving systems by selecting these systems using their proper motion.  These two effects combined likely produce the observed large vertical velocity dispersion.

\subsection{Magnetic Activity}
\label{sec:magact}
\begin{figure}
\figurenum{6}
\includegraphics[angle= 0,width={\columnwidth}]{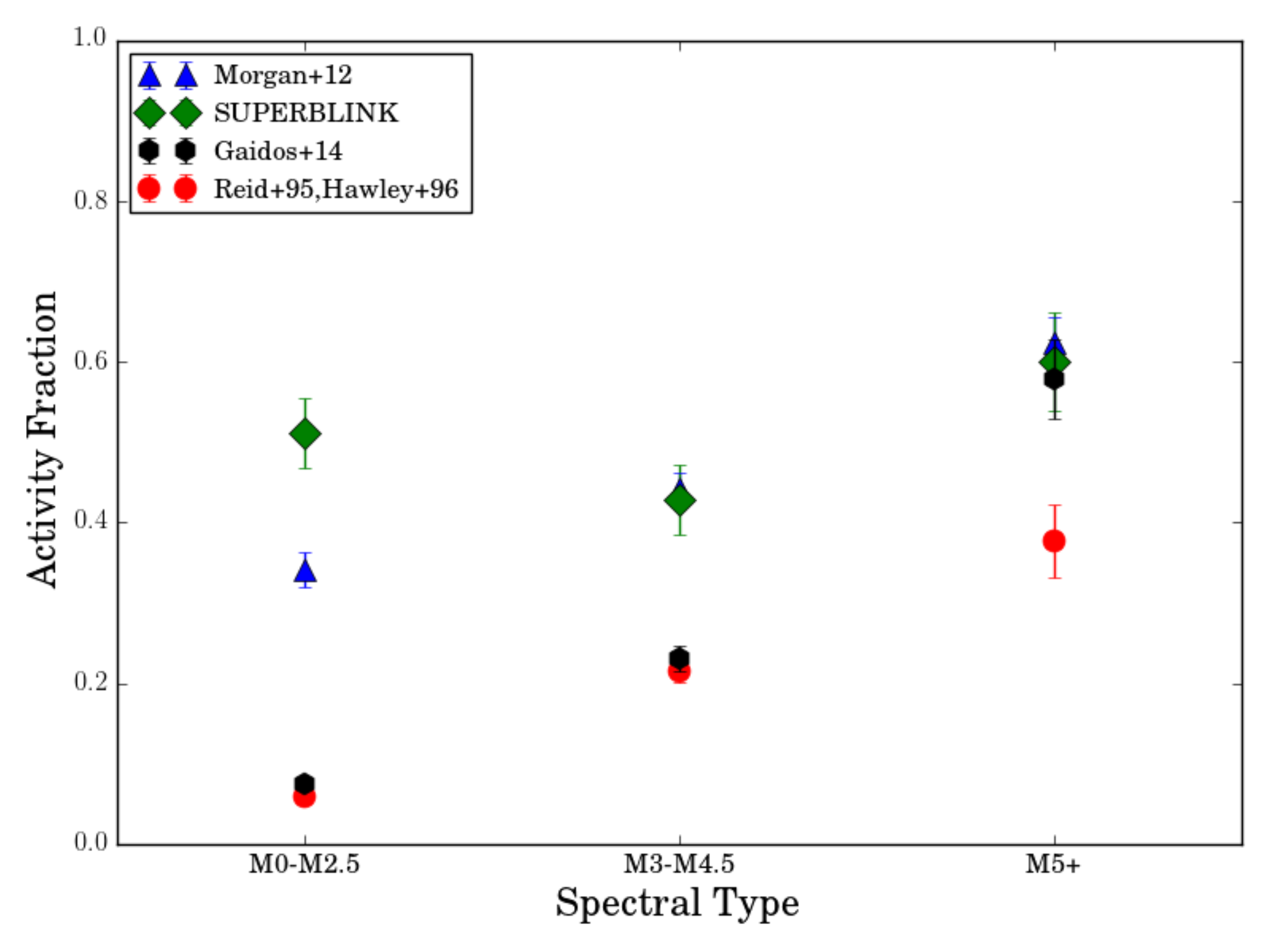}
\caption{The fraction of active stars as a function of spectral type is shown for the full SUPERBLINK sample of WD+dMs (green diamonds), the WD+dMs from \citet{Morgan12} (blue triangles) and the samples of nearby single M dwarfs from \citet{Gaidos14} (black hexagons) and the PMSU survey (\citealt{Hawley96}; gray circles).  Errors were drawn from the binomial distribution for each sample.  The SUPERBLINK WD+dMs and the SDSS WD+dMs show high activity fractions compared to the single M dwarfs at spectral types earlier than M5.  At late types, the difference between activity fractions for single stars and WD+dMs decreases or disappears entirely.
}
\label{fig:afcomp}
\end{figure}

Using H$\alpha$ chromospheric emission as a tracer for magnetic activity, we classified each M dwarf as active or inactive.  We integrated over a 12~\AA~region centered on H$\alpha$, with the continuum set by measuring the flux in 5~\AA~regions on either side of the 12~\AA\ region \citep{West04}.  A star is classified as active if the equivalent width measured is $>$ 1~\AA.  We visually examined each of the M dwarf component spectra to verify the activity classification.  The automatic identification of active systems failed in a number of cases with clear, narrow H$\alpha$ line emission.  Since these systems are close binaries, it is possible that the H$\alpha$ line moved out of the integrated region.  In general, we relied on the visual determination for our analysis.  Overall, 55\% of our sample was classified as active. Figure \ref{fig:afcomp} shows the activity fraction as a function of spectral type for our sample, as well as WD+dMs from \citet{Morgan12}, single M dwarfs from SUPERBLINK \citep{Gaidos14}, and single M dwarfs from the Palomar-Michigan State University (PMSU) survey of nearby M dwarfs \citep{Reid95,Hawley96}.  We chose spectral type bins representing early, mid and late type M dwarfs. The binary sample from SUPERBLINK, as well as the \citet{Morgan12} sample show higher activity fractions at early and mid-spectral types compared with the samples of single M dwarfs.  Using the observed activity fractions for M dwarfs from SDSS DR5 and a simple 1D dynamical model, \citet{West08} showed that M0-M2 dwarfs have activity lifetimes ranging from 0.8-1.2 Gyr, mid-M dwarfs (M3-M4) have activity lifetimes 2-4 Gyr, and late M dwarfs (M5-M6) have activity lifetimes 7-8 Gyr.  For these binaries with early or mid-type M dwarfs, the high activity fractions compared to single M dwarf samples indicates that the companion likely plays an important role in lengthening the activity lifetime.  For systems with late-type M dwarf components, both binary samples show higher activity fractions than the PMSU survey, but are the same as the \citet{Gaidos14} sample.  Also selected from SUPERBLINK, the M dwarfs from \citet{Gaidos14} however all have \textit{J} $<$ 9; this magnitude limit may introduce a selection bias toward younger M dwarfs that would have the largest effect for the latest M spectral types.

\begin{figure}
\figurenum{7}
\includegraphics[angle= 0,width={\columnwidth}]{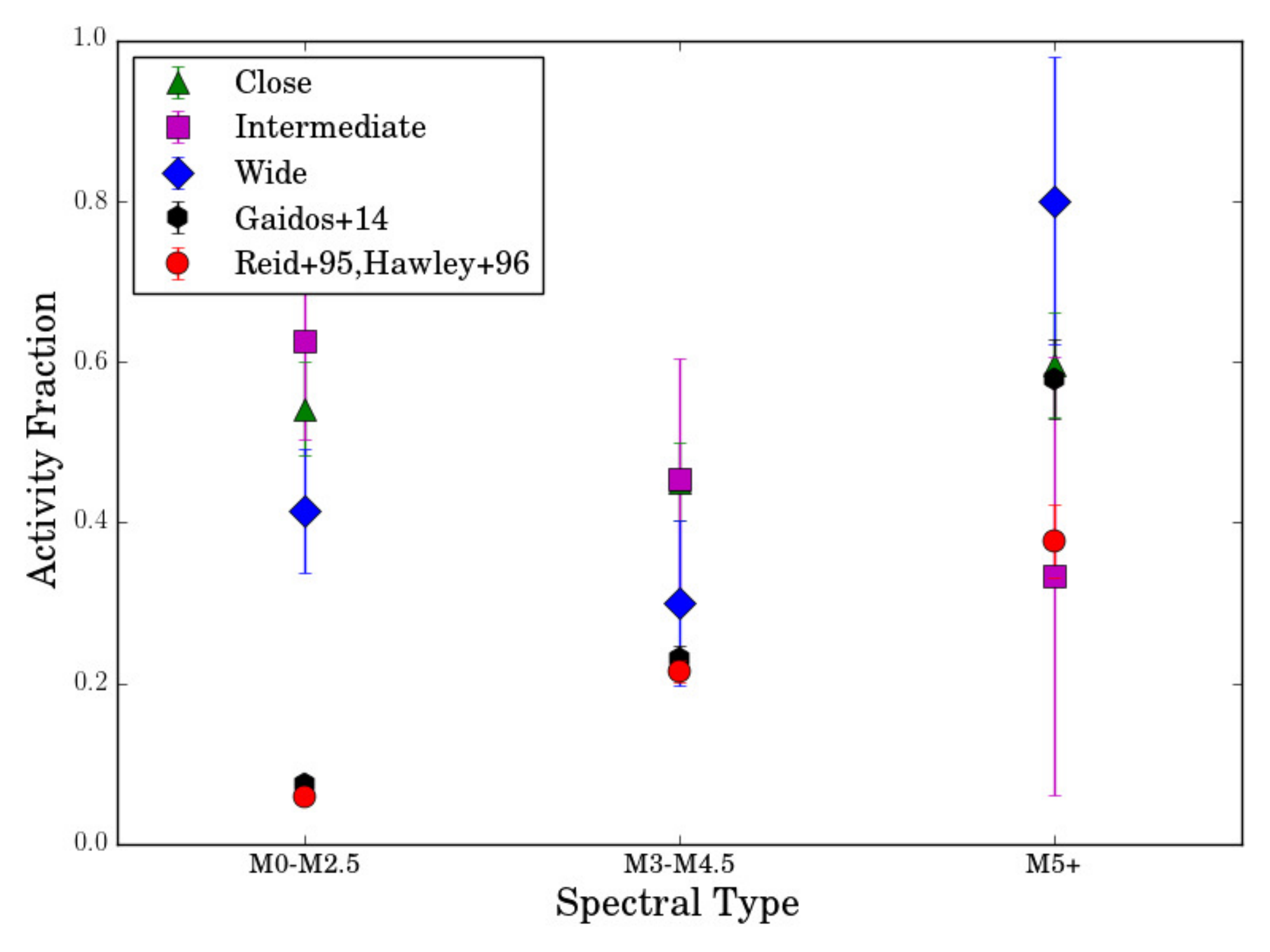}
\caption{Activity fractions for the SUPERBLINK WD+dMs are shown based on our estimated projected binary separations and compared to the same single M dwarf samples as Figure \ref{fig:afcomp}.  Errors were again drawn from the binomial distribution.  Separation bins were chosen based on definitions given by\citet{Willems04}.  Close separations (a $\leq$ 0.88 AU) are shown as green triangles, intermediate separation (0.88 AU < a $\leq$ 2.57 AU) are shown as red squares, and wider systems (a > 2.57 AU) are shown as blue diamonds.  For early-type systems, WD+dMs at all separations have higher activity fractions than the samples of single stars.  For mid-types, the activity fraction for the widest system matches that of the sample of single M dwarfs hinting that there may be a threshold where the binary companion plays a less significant role.   At late types, we have the fewest objects, especially at intermediate and wide separations.  More data are needed at late-types to place constraints on the relationship between binary separation and active fraction.
}
\label{fig:afsep}
\end{figure}

We investigated the effects that binary separation could have on the activity fraction in our sample as compared to single stars. Figure \ref{fig:afsep} shows the activity fractions for our sample binned by their likelihood of having gone through common envelope evolution.  For this, we used the orbital period bins derived for white dwarf main sequence binaries by \citet{Willems04}.  For early-type M dwarfs in WD+dMs, we continue to see high activity fractions across all binary separations.  For mid-type WD+dMs with wider separations, the activity fraction for WD+dMs matches that of the single M dwarf populations.  We note that the relatively small numbers in our sample do not allow us to confidently identify a separation threshold at which the binary companion does not significantly affect the activity lifetime.  However, the possibility that such a threshold may exist warrants further investigation.  \citet{Morgan12} found a similar trend for widely separated WD+dMs at mid- and late-M spectral types. Though the majority of our sample and the \citet{Morgan12} sample are made up of close binaries, magnetic activity in the widest binaries may be affected by a very close tertiary companion \citep{Makarov08}.    At closer separations, we continue to see high activity fractions indicating longer activity lifetimes for mid-type M dwarfs with close companions.  At late spectral types, we see high activity fractions at all projected binary separations.  Our sample contains relatively few late-type M dwarfs with WD companions, so further investigation is needed to understand how binary separation may affect the activity lifetimes of late-type M dwarfs with binary companions.  

Since WD+dM systems contain the remnant of an evolved star, the M dwarfs in these systems are likely older than the complementary samples of nearby single M dwarfs and M dwarfs with main sequence companions (see also Section \ref{sec:kinematics}).  Using the initial-final mass function for WDs from \citet{Kalirai08} combined with our median WD mass (0.62 M$_{\sun}$), we estimated that the most likely mass of a white dwarf progenitor for our sample of WD+dMs is 2.07 M$_{\sun}$.  Typically, 2 M$_{\sun}$ stars have main sequence lifetimes of $\sim$1 Gyr \citep{Paxton10}.  Comparing this to the activity lifetimes derived by \citet{West08}, most of the M0-M2 dwarfs in WD+dMs should be older than what their single M dwarf activity lifetimes might suggest (i.e. their single star counterparts of similar age should be mostly inactive).  This is further evidence that a close companion prolongs the activity lifetimes in the stars in a significant way.  As outlined in section \ref{sec:intro}, tidal effects play a role in keeping the M dwarfs in the closest binaries rotating quickly.  In particular, \citet{RebassaMansergas13a} showed that all of the M dwarfs in a sample of confirmed post-common envelope binaries were active.  We also see high activity fractions for binaries at intermediate separations.  The physical mechanism behind this is less clear, but may be related to early inhibition of angular momentum loss through disk disruption or dissipation.

\section{Conclusion}
\label{sec:conclusion}
We have identified and characterized WD+dMs selected from the SUPERBLINK proper motion survey.  It is the first catalog of WD+dMs for which all objects have well-measured proper motions, enabling a first look at the Galactic kinematics of these objects.  We used the presence of H$\alpha$ to investigate the effects of a close WD companion on M dwarf magnetic activity fractions.  We compared the M dwarf activity fractions for the SUPERBLINK WD+dMs and SDSS DR8 WD+dMs to nearby samples of single M dwarfs and examined how activity fraction changes with projected binary separation.  We summarize our findings below: 
\begin{enumerate}
\item We used a UV-Optical-IR color selection to select for WD+dMs in the SUPERBLINK proper motion survey.  Spectroscopic follow-up at MDM Observatory yielded 178 previously uncharacterized systems.  Combined with previously identified binaries, we now have confirmed 470 nearby WD+dMs in the SUPERBLINK proper motion survey to date.
\item We fit each of our spectra with WD models and M dwarf templates to derive the physical parameters of each of the stars in the WD+dM binaries.  Output parameters of these fits include WD effective temperature and gravity, M dwarf spectral type, lower limits on the radial velocities, approximate binary separations, magnetic activity (traced by H$\alpha$).  Combined with synthetic absolute NUV photometry for the WDs, we calculated a spectroscopic distance to the WD in each system.  Using a SpT-M$_{J}$ relation, we calculate a spectroscopic distance to the M dwarf in each system.  We found that the SUPERBLINK sample is on average significantly closer than previous samples of WD+dMs identified from SDSS.
\item Combining the proper motions from SUPERBLINK with each WD+dM position, estimated radial velocity, and M dwarf spectroscopic distance, we calculated 3D space velocities.  We found evidence for asymmetric drift in the binaries containing inactive M dwarfs, indicating this subsample is likely drawn from an older population than the systems with active M dwarfs.  We also found that the entire sample has a vertical (W) velocity dispersion larger than that of local field stars, suggesting the entire sample is drawn from an older population of stars than the complementary high proper motion samples of single M dwarfs.
\item The magnetic activity fraction for M dwarfs in the SUPERBLINK WD+dM sample corroborates previous findings that the activity fraction in these binaries is higher for early- and mid-M dwarfs than for samples of single M dwarfs.  At late spectral types, this difference in activity fraction decreases or disappears.
\item We examined how activity fraction changes as a function of binary separation and found enticing evidence of a limiting separation where the existence of a companion does not affect the activity fraction of a sample of mid-M dwarfs.  For early spectral types, we found high activity fractions at each of our separation bins.  For mid-spectral types, we found high activity fractions for systems with close and intermediate separations.  At wide separations, however, activity fraction for the binaries matched the single star populations suggesting the possibility of a separation threshold where companions do not contribute to higher activity fraction.    
\end{enumerate}

This work, combined with the evidence from previous studies of WD+dMs, highlights the need to define three distinct separation regimes for understanding how M dwarfs (and stars in general) are affected by evolved companions.  Categories of binaries are usually restricted to ``close", where interactions dominate and change the evolution of the two stars and ``wide", where each component is assumed to evolve as a single star.  However, we argue that an intermediate separation category should be used, as we clearly find evidence for higher activity fractions at projected separations outside the regime of tidal interaction.  Higher activity fractions indicate longer activity lifetimes and therefore longer spindown times.  To achieve this, there must be some inhibition of angular momentum loss either at early epochs via disk disruption or dissipation or throughout the binary lifetime via magnetic field locking.  \citet{Meibom07} interpreted shorter binary primary rotation periods for solar-type stars in a young open cluster (M35) as the result of early disk disruption or dissipation.  Disk disruption or dissipation may cause binaries to follow a delayed Skumanich-type \citep{Skumanich72} power law, where a star in a binary begins its main sequence life with a higher initial rotation rate compared to a single star, and therefore takes longer to spindown.  A difference in the normalization of the Skumanich-type relation has been argued previously in the more extreme case of cataclysmic variables \citep{Patterson84, Knigge11}.  It is also possible that the M dwarf spindown is altered by the presence of the white dwarf magnetic field.  \citet{Cohen12} showed that the extreme case of a magnetic cataclysmic variable (CV), the presence of a strong white dwarf magnetic field could reduce the M dwarf outflow by reducing open field regions.  However, despite the common occurrence of magnetic white dwarfs in interacting binaries like CVs, there are no magnetic white dwarfs known in non-interacting WD+dMs \citep{Holberg16, Liebert05}.  Any alteration of M dwarf spindown due to magnetic field interactions with the white dwarf would have to take place in the context of low WD magnetic field strengths. More work is needed to understand how binaries outside the regime of tidal interactions may be affected by magnetic field interactions.

Comparison to other binaries containing M dwarfs and other main sequence stars is needed to fully understand the effects of the white dwarf progenitor's evolution.  However, WD+dMs remain useful probes of how close companions affect the magnetic activity of M dwarfs, largely due to their bright and spectroscopically distinct nature.  

\acknowledgments
The authors would like to thank Shaun Akhtar, Tim Scanlon, Brad Nelson, and Jules Halpern for assistance obtaining observations for this project.  A.A.W. and J.N.S. acknowledge funding from the United States National Science Foundation (NSF) grants AST-1109273 and AST-1255568 and the support from the Research Corporation for Science Advancement's Cottrell Scholarship.  J.N.S. would like to thank Eunkyu Han for assistance with \texttt{MESA}.  This research was made possible through the GALEX Guest Investigator program under NASA grant NNX09AF88G.  SL also acknowledges support from the NSF under the astronomy research grant AST-0908406.  The authors also thank the anonymous referee for thoughtful comments that improved the quality of the manuscript.  This research made use of Astropy, a community-developed core Python package for Astronomy \citep{astropy}.  Figures were created using the Python-based plotting library \texttt{matplotlib} \citep{Hunter07}.

\bibliography{ms}
\begin{rotatetable*}
\begin{deluxetable*}{lccccccccccccc}
\tablenum{1}
\tablecaption{Astrometric and Photometric Properties of the WD+dMs in the SUPERBLINK proper motion survey  \label{tab:astrophot}}
\tablehead{
\colhead{Name} &
 \colhead{RA} &
  \colhead{Dec} &
   \colhead{[$\mu_{\alpha}, \mu_{\delta}$]} &
    \colhead{FUV} & 
    \colhead{NUV} & 
    \colhead{V} & 
    \colhead{J} &
   \colhead{H} &
    \colhead{K$_{s}$} &
     \colhead{NUV-V} & 
     \colhead{V-K$_{s}$} &
      \colhead{Type\tablenotemark{a}} & 
      \colhead{Reference} \\
 & 
 \colhead{(J2000.0)} & 
 \colhead{(J2000.0)} & 
 \colhead{(mas yr$^{-1}$)} & 
 \colhead{mag} & 
 \colhead{mag} & 
 \colhead{mag} &
  \colhead{mag} &
\colhead{mag} & 
\colhead{mag} & 
& 
&
 & 
}
\startdata
  PM I00010+3527 & 0.274231 & 35.461109 & [46.0,1.0] & 18.97 & 19.05 & 17.96 & 15.17 & 14.58 & 14.25 & 1.09 & 3.71 & DA/M & This Work\\
  PM I00057+2915 & 1.445243 & 29.260513 & [-51.0,-58.0] & $\cdots$ & $\cdots$ & 18.31 & 16.78 & 15.41 & 15.34 & $\cdots$ & 2.97 & DA/M: & RM12\\
  PM I00090+4348 & 2.268687 & 43.81242 & [70.0,-35.0] & 18.57 & 17.92 & 16.77 & 13.69 & 13.05 & 12.81 & 1.15 & 3.96 & DA/M & This Work\\
  PM I00136+0019 & 3.413117 & 0.323517 & [406.0,-188.0] & 19.70 & 16.42 & 15.62 & 15.15 & 15.21 & 15.1 & 0.8 & 0.52 & (WD/M) & RM12\\
  PM I00139-1108E & 3.497771 & -11.144092 & [24.0,32.0] & 17.50 & 17.79 & 16.12 & 13.49 & 12.85 & 12.7 & 1.67 & 3.42 & DA/M & RM12,M12\\
  PM I00219-1103 & 5.491246 & -11.058861 & [-6.0,-44.0] & 21.95 & 19.71 & 18.15 & 14.84 & 14.37 & 14.03 & 1.56 & 4.12 & DA/M & RM12,M12\\
  PM I00241+0548 & 6.030123 & 5.815727 & [134.0,-30.0] & 18.34 & 17.34 & 16.26 & 12.73 & 12.15 & 11.88 & 1.08 & 4.38 & DA/M & This Work\\
  PM I00263+1444 & 6.585003 & 14.735963 & [22.0,-34.0] & 17.14 & 17.31 & 17.15 & 14.68 & 14.16 & 13.92 & 0.16 & 3.23 & DA/M & RM12,M12\\
  PM I00278-0010 & 6.958331 & -0.17318 & [63.0,-11.0] & 20.00 & 18.73 & 18.04 & 15.01 & 14.65 & 14.29 & 0.69 & 3.75 & DB/M & RM12,M12\\
  PM I00305+0706 & 7.637668 & 7.116136 & [35.0,-51.0] & 18.13 & 17.86 & 17.93 & $\cdots$ & $\cdots$ & $\cdots$ & -0.07 & $\cdots$ & (DA/M) & RM12\\
  PM I00336+0041 & 8.402044 & 0.697649 & [-41.0,-61.0] & $\cdots$ & $\cdots$ & 19.39 & 15.11 & 14.41 & 14.15 & $\cdots$ & 5.24 & WD/M & RM12\\
  PM I00342+0004 & 8.555728 & 0.07724 & [75.0,-17.0] & $\cdots$ & $\cdots$ & 17.82 & 15.06 & 14.54 & 14.22 & $\cdots$ & 3.6 & WD/M & RM12\\
  PM I00365+1118 & 9.145305 & 11.302482 & [-55.0,-23.0] & 17.44 & 17.42 & 17.86 & 16.66 & 15.87 & 16.38 & -0.44 & 1.48 & (WD/M) & This Work\\
  PM I00380+0834 & 9.51839 & 8.571456 & [2.0,-55.0] & 23.07 & 20.29 & 17.83 & 13.98 & 13.46 & 13.2 & 2.46 & 4.63 & DA/M & RM12,M12\\
  PM I00389+0106 & 9.736704 & 1.114177 & [-29.0,-46.0] & $\cdots$ & $\cdots$ & 18.65 & 14.67 & 14.01 & 13.71 & $\cdots$ & 4.94 & WD/M & RM12,M12\\
\enddata
\tablenotetext{a}{Types of WD+dMs are given according to the convention outlined in \citet{RebassaMansergas10}.  For objects found in both SDSS and SUPERBLINK, we have used the types given by the catalogue maintained at \url{http://sdss-wdms.org/}.  A colon after a component type indicates the classification of that component is uncertain.  Parentheses indicate that the system is a WD+dM candidate. More details about these classifications can be found in \citet{RebassaMansergas10}.}
\tablerefs{F05 - \citet{Farihi05}, F06 - \citet{Farihi06}, H07 - \citet{Hoard07}, L82 - \citet{Lanning82}, L14 - \citet{Li14},
	L13 - \citet{Limoges13}, M76 - \citet{Margon76}, M12 - \citet{Morgan12}, O10 - \citet{Ostensen10}, RM12 - \citet{RebassaMansergas12}, 
	R14 - \citet{Ren14}, T07 - \citet{Tappert07}, W03 - \citet{Wachter03}}
\tablecomments{(This table is available in its entirety in machine-readable format.)}
\end{deluxetable*}
\end{rotatetable*}

\movetabledown=1in
\begin{rotatetable*}
\begin{deluxetable*}{lccccccccccccc}
%\rotate
\tabletypesize{\footnotesize}
\tablenum{2}
%\tablecolumns{14}
\tablecaption{Spectral Types, Radial Velocities, and Spectroscopic Distances of the newly identified SUPERBLINK WD+dMs \label{tab:rvsspt}}
\tablehead{
\colhead{Name} &
\colhead{dM} &
\colhead{WD T$_{eff}$} &
\colhead{WD $log(g)$} &
\colhead{WD RV} &
\colhead{$\sigma_{WDRV}$} &
\colhead{dM RV} &
\colhead{$\sigma_{dMRV}$} &
\colhead{System RV} &
\colhead{$\sigma_{sysRV}$} &
\colhead{Sep.} &
\colhead{Activity Flag} &
\colhead{dM Dist.} &
\colhead{WD Dist.}
\\
 & 
 \colhead{SpT} &
 \colhead{(K)} &
 &
\colhead{(km s$^{-1}$)} &
\colhead{(km s$^{-1}$)} &
\colhead{(km s$^{-1}$)} &
\colhead{(km s$^{-1}$)} &
\colhead{(km s$^{-1}$)} &
\colhead{(km s$^{-1}$)} &
 \colhead{(AU)} &
 &
 \colhead{(pc)} &
 \colhead{(pc)}
 }
 \startdata
  PM I00090+4348 & 3.0 & 14000.0 & 7.75 & 159.9 & 59.4 & -207.4 & 218.3 & -2.8 & 1.9 & .33 & 1 & 197. & 220.\\
  PM I00139-1108E & 0.0 & 25000.0 & 8.5 & 177.6 & 172.2 & -139.4 & 69.5 & 178.1 & 69.3 & .94 & 0 & 354. & $\cdots$\\
  PM I00219-1103 & 4.5 & 10000.0 & 8.25 & 120.7 & 75.2 & -334.5 & 178.8 & 143.1 & 57.4 & .02 & 1 & 121. & $\cdots$\\
  PM I00241+0548 & 2.0 & 16000.0 & 7.5 & -37.2 & 112.4 & 35.1 & 52.8 & 30.3 & 31.3 & 155.40 & 1 & 206. & 251.\\
  PM I00263+1444 & 4.5 & 20000.0 & 8.0 & 43.0 & 141.8 & -119.2 & 84.6 & 120.6 & 84.0 & .12 & 1 & 101. & $\cdots$\\
  PM I00278-0010 & 4.0 & 12000.0 & 7.0 & 49.6 & 151.0 & -116.8 & 87.2 & 124.4 & 86.8 & .17 & 0 & 162. & $\cdots$\\
  PM I00380+0834 & 3.0 & 10000.0 & 7.75 & -79.7 & 265.0 & 104.2 & 130.8 & -60.0 & 130.7 & 1.23 & 1 & 155. & $\cdots$\\
  PM I00389+0106 & 4.0 & 25000.0 & 6.0 & 150.3 & 276.7 & -354.0 & 159.1 & 359.5 & 159.0 & .02 & 0 & 141. & $\cdots$\\
  PM I00532-0032 & 1.5 & 25000.0 & 8.0 & 166.8 & 175.2 & -168.9 & 78.8 & 180.0 & 78.8 & 210.82 & 0 & 247. & $\cdots$\\
  PM I00554-0831 & 6.5 & 14000.0 & 8.0 & 64.9 & 53.0 & 58.6 & 576.3 & -102.9 & 34.8 & .07 & 0 & 79. & 114.\\
  PM I01101+1326 & 5.0 & 25000.0 & 7.5 & 8.7 & 121.4 & -29.2 & 90.2 & 172.2 & 74.7 & 1.16 & 0 & 96. & $\cdots$\\
  PM I01103-1853E & 3.5 & 10000.0 & 7.75 & -62.5 & 93.1 & 83.9 & 41.9 & -30.4 & .3 & 1.62 & 0 & 177. & 163.\\
  PM I01106-0108 & 0.0 & 50000.0 & 7.75 & 88.2 & 70.8 & -69.2 & 30.3 & 148.0 & 28.9 & 2.84 & 1 & 346. & $\cdots$\\
  PM I01163-0436 & 2.5 & 35000.0 & 8.0 & -52.1 & 126.1 & -26.4 & 32.3 & -7.0 & 25.2 & .15 & 0 & 487. & 448.\\
  PM I01187+0505 & 2.5 & 14000.0 & 7.75 & 67.4 & 98.6 & -175.2 & 56.4 & 57.7 & 52.2 & .07 & 1 & 310. & 197.\\
\enddata
\tablecomments{(This table is available in its entirety in machine-readable format.)}
\end{deluxetable*}
\end{rotatetable*}

\begin{deluxetable*}{lcccccc}
\tabletypesize{\normalsize}
\tablenum{3}
\tablecolumns{7}
\tablecaption{3D Space Velocities for SUPERBLINK WD+dMs \label{tab:kinematics}}
\tablehead{
\colhead{Name} &
\colhead{$U$} &
\colhead{$\sigma_{U}$} &
\colhead{$V$} &
\colhead{$\sigma_{V}$} &
\colhead{$W$} &
\colhead{$\sigma_{W}$} 
\\
&
\colhead{(km s$^{-1}$)} &
\colhead{(km s$^{-1}$)} &
\colhead{(km s$^{-1}$)} &
\colhead{(km s$^{-1}$)} &
\colhead{(km s$^{-1}$)} &
\colhead{(km s$^{-1}$)}
}
\startdata
  PM I00090+4348 & 39.73 & 6.63 & -18.04 & 5.22 & -30.1 & 7.64\\
  PM I00139-1108E & 73.23 & 14.62 & 86.46 & 23.95 & -152.78 & 65.93\\
  PM I00219-1103 & 4.15 & 4.45 & 28.29 & 17.61 & -135.95 & 54.76\\
  PM I00241+0548 & 64.45 & 10.81 & -37.53 & 19.51 & -47.58 & 26.59\\
  PM I00263+1444 & 33.03 & 23.1 & 69.56 & 51.76 & -93.91 & 62.18\\
  PM I00278-0010 & 67.34 & 16.43 & 28.67 & 37.9 & -111.4 & 76.99\\
  PM I00380+0834 & -28.85 & 35.48 & -51.12 & 68.22 & 32.13 & 106.03\\
  PM I00389+0106 & 54.72 & 33.8 & 145.89 & 68.01 & -322.63 & 139.85\\
  PM I00532-0032 & -16.1 & 20.7 & 42.16 & 32.2 & -186.05 & 70.66\\
  PM I00554-0831 & 10.84 & 7.44 & -30.36 & 9.53 & 104.2 & 32.99\\
  PM I01101+1326 & 97.82 & 31.54 & 73.04 & 37.56 & -126.84 & 56.56\\
  PM I01103-1853E & -33.54 & 7.66 & -15.81 & 6.11 & 29.46 & 1.43\\
  PM I01106-0108 & 29.56 & 13.11 & 58.45 & 12.79 & -132.27 & 26.23\\
  PM I01163-0436 & -70.28 & 18.8 & -6.86 & 14.45 & -16.95 & 24.2\\
  PM I01187+0505 & 103.78 & 24.06 & -3.9 & 22.04 & -26.01 & 44.1\\
  \enddata
\tablecomments{(This table is available in its entirety in machine-readable format.)}
\end{deluxetable*}

\begin{deluxetable*}{lcccccc}
\tablenum{4}
\tablecolumns{7}
\tablecaption{Population Kinematics \label{tab:popkinematics}}
\tablehead{
\colhead{Pop. ID} &
\colhead{$\mu_{u}$} &
\colhead{$\mu_{v}$} &
\colhead{$\mu_{w}$} &
\colhead{$\sigma_{u}$} &
\colhead{$\sigma_{v}$} &
\colhead{$\sigma_{w}$} 
}
\startdata
All & 18.9$^{3.8}_{3.8}$ & -15.7$^{2.8}_{2.8}$ & 17.2$^{4.4}_{4.4}$ & 73.6$^{3.2}_{3.4}$ & 54.1$^{2.5}_{2.7}$ & 82.3$^{3.8}_{4.0}$\\
Active & 19.4$^{4.4}_{4.4}$ & -10.1$^{3.3}_{3.3}$ & 17.9$^{6.0}_{6.1}$ & 60.5$^{3.4}_{3.8}$ & 44.2$^{2.8}_{3.0}$ & 84.8$^{5.0}_{5.3}$\\
Inactive & 17.8$^{6.1}_{6.0}$ & -22.1$^{4.6}_{4.6}$ & 16.6$^{6.0}_{6.1}$ & 88.4$^{5.1}_{5.4}$ & 64.6$^{4.0}_{4.4}$ & 78.7$^{5.5}_{5.8}$\\
\enddata
\label{tab:kinematics}
\end{deluxetable*}

\end{document}